\documentclass[aps,prl,twocolumn,superscriptaddress]{revtex4-2}

\newcommand{\eq}[1]{\begin{equation}#1\end{equation}}
\newcommand{\be}{\begin{equation}}
\newcommand{\ee}{\end{equation}}
\newcommand{\dd}{\mathrm{d}}
\newcommand{\e}{\mathrm{e}}

\newcommand{\Tr}{\mathrm{Tr\,}}

\usepackage{amsmath}
\usepackage{amsfonts}
\usepackage{xcolor}
\usepackage{braket}
\usepackage{bbold}
\usepackage{graphics}
\usepackage{graphicx}
\usepackage{comment}
\usepackage{leftidx}
\usepackage{psfrag}
\usepackage{color}
\usepackage{colordvi}
\usepackage{bbm}
\usepackage{hyperref}
\hypersetup{colorlinks,bookmarksopen,bookmarksnumbered,
 citecolor=red,
 linkcolor=blue,
 pdfstartview=green,
 urlcolor=magenta}
 \usepackage{soul}

\begin{document}

\title{Measurement-induced entanglement Hamiltonian}

\author{Viktor Eisler}
\affiliation{Institute of Physics, University of Graz, Universit\"atsplatz 5, A-8010 Graz, Austria}
\author{Erik Tonni}
\affiliation{SISSA and INFN Sezione di Trieste, via Bonomea 265, I-34136 Trieste, Italy}

\begin{abstract}
We study the entanglement Hamiltonian of an infinite hopping chain in its ground state, after partial projective measurements in the occupation basis. For a segment separated by two measurement regions from the rest of the chain, we show that the reduced density matrix can be related to a grand-canonical state via a conformal mapping and a gauge transformation in the underlying field-theory description. The entanglement Hamiltonian is then described by a local inverse temperature that vanishes as a square root around the endpoints and is independent of the particular measurement outcome. In sharp contrast, the local chemical potential is shown to be related to the induced charge density in the segment. Hence the entanglement Hamiltonian of a post-selected state contains much more information on the measurement outcome than the respective entropy.

\end{abstract}

\maketitle

Measurements and entanglement both play a key role in quantum computation 
\cite{NielsenChuang}, and their subtle interplay have been exploited as a key ingredient in protocols like teleportation \cite{Bennett:1992tv}, error-correction \cite{Shor:1995hbe}
and measurement-based quantum computation
\cite{Raussendorf:2001zim, Raussendorf:2003rug,Briegel:2009inn}.
Entanglement properties also proved to be invaluable in the study of quantum many-body systems, with various tools developed during the past few decades \cite{Calabrese:2009kka, Laflorencie:2015eck,Casini:2022rlv}. 
In contrast, the role of measurements in developing novel phases of quantum matter is much less understood. Recent years have witnessed enormous progress in several directions, such as uncovering measurement-induced phase transitions 
\cite{Skinner:2018tjl, Li:2018mcv,Li:2019zju,Jian:2019mny,Bao:2019qah,Potter:2021lxx,Noel:2021hez,Koh:2022ajm,Hoke:2023nmt}, and various mechanisms inducing or altering criticality 
\cite{Ashida:2016liw,Buchhold:2021par,Minoguchi:2021igh,
Agrawal:2021ukw,Garratt:2022ycp,Murciano:2023nqr,Tang:2024wlh}
and long-range entanglement 
\cite{Tantivasadakarn:2021vel,
Verresen:2021wdv,Lu:2022jax,Iqbal:2023wvm,Foss-Feig:2023uew,Zhu:2022bpk,
Chan:2024tdk,Milekhin:2024imt}.

Within this arena, a particularly interesting question is how \emph{partial} projective measurements in the ground state modify the structure of entanglement between the remaining, unmeasured parts of the system. For quantum critical systems in one spatial dimension, this setup can be addressed via tools of conformal field theory (CFT) \cite{DiFrancesco:1997nk}. Focusing on particular post-selected measurement outcomes, the entanglement entropy between two regions separated by the measured sites in a critical quantum chain was obtained in \cite{Rajabpour:2015uqa, Rajabpour:2015xkj, Najafi:2016kwb}, using methods of boundary CFT \cite{Cardy:1984bb,Cardy:1989ir,Cardy:2004hm,Cardy:2016fqc}.
Similar methods were applied to study entanglement swapping between two copies of the chain, induced by partial Bell-state measurements \cite{Hoshino:2024kxk}. However, instead of post-selection, a more useful quantifier is the \emph{averaged} entropy over all possible outcomes, weighted by their Born probabilities. This measurement-induced entropy is related to localizable entanglement \cite{Verstraete:2003spp,Popp:2004yt}, has relevance for the sign problem \cite{Lin:2022jvx}, as well as other operational meanings \cite{McGinley:2024xaw}.
Remarkably, numerical studies revealed that it also has universal features in critical chains \cite{Lin:2022jvx,Cheng:2023bvw}. Very recently, an analytical treatment for the Tomonaga-Luttinger liquid has shown that the result can indeed be written as a Born-average over appropriate conformal boundary conditions \cite{Khanna:2025fac, Khanna:2025mrz}.

Motivated by a better understanding of this result,
in this paper we investigate the measurement-induced entanglement Hamiltonian (EH), focusing on a lattice realization of the massless Dirac field
under partial projective measurements of the fermion density. Remarkably, we find that the EH remains local for any post-selected state that corresponds to a measurement outcome flowing towards a conformal boundary condition. However, besides the local inverse temperature commonly found for many-body ground states \cite{Dalmonte:2022rlo}, the EH also acquires an inhomogeneous chemical potential. This reflects a nontrivial charge density within the subsystem, induced by the combined effects of projective measurements and charge conservation.
The measurement-induced entanglement has thus a purely thermodynamic interpretation, given by a Born-average of entropies as a function of the total induced charge. Moreover, the EH retains additional information on each particular outcome via the induced density, encoded in the chemical potential. Our results are nicely confirmed by lattice calculations.


%
\begin{figure}[t]
\center
\includegraphics[width=\columnwidth]{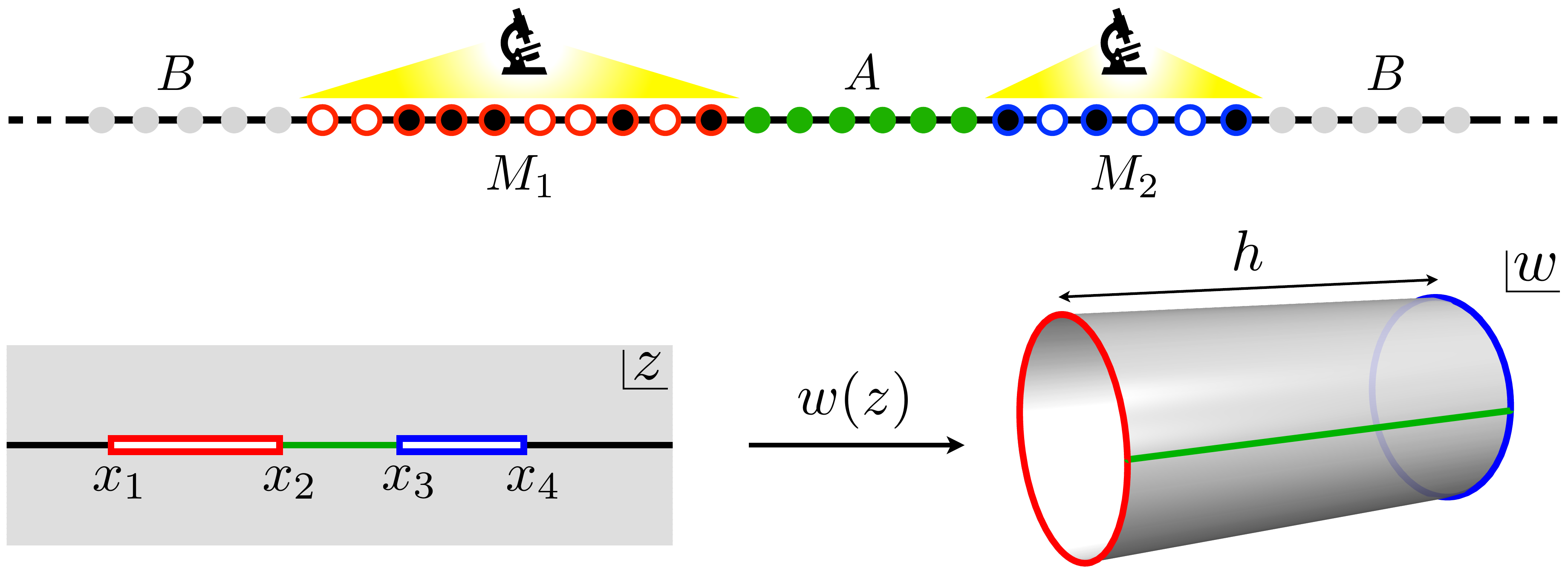}
\caption{Measurement setup (top) and conformal map from the compactified complex plane  
with two slits to the cylinder (bottom). 
Filled and empty circles in the segments $M_1$ and $M_2$ 
represent the measurement outcomes $1$ and $0$ respectively.}
\label{fig:map}
\end{figure}
%

Consider the ground state of an infinite chain of non-interacting fermions described by the Hamiltonian
%
\begin{equation}
\hat H= -\frac 1 2 \sum_{n=-\infty}^{\infty} (c^{\dag}_n c_{n+1} + c^{\dag}_{n+1} c_{n}) \, .
\label{H}
\end{equation}
Local projective measurements are performed in the occupation number basis, in two disconnected segments $M_1$ and $M_2$
(see the top part of Fig.~\ref{fig:map}).
The unnormalized post-measurement state $\hat\rho_\mathbf{m}$ corresponds to the outcome $\mathbf{m}$ and occurs with Born probability $p_\mathbf{m}=\Tr \hat\rho_\mathbf{m}$. We are interested in the entanglement between the remaining parts $A$ and $B$ of the chain, that are separated by the measurement regions. This is characterized by the reduced density matrix (RDM) $\hat\rho_{\mathbf{m},A}=\mathrm{Tr}_B \, \hat\rho_\mathbf{m}/\Tr \hat\rho_\mathbf{m}$,
which can be written in terms of the EH as $\hat\rho_{\mathbf{m},A}\propto\e^{-\mathcal{H}_{\mathbf{m}}}$.
%


The expression of the EH can be found by considering the low-energy description of the problem via a massless Dirac field. As illustrated on the bottom left panel of Fig.~\ref{fig:map}, the RDM $\hat\rho_{\mathbf{m},A}$ is represented by a path integral on the complex plane $z$, with a cut along $A$ (green) as well as two additional slits along the measurement intervals $M_1=(x_1, x_2)$ (red) and $M_2=(x_3, x_4)$ (blue).
Indeed, along these slits the fermion fields are constrained to values determined by the coarse-grained measurement outcomes.
Remarkably, there exists a conformal mapping $w(z)$ that transforms this rather complicated geometry onto a cylindrical surface \cite{Nehari:book} (see the bottom part of Fig.~\ref{fig:map}), which is key to the calculation of the entropy \cite{Rajabpour:2015xkj,Khanna:2025fac}. In particular, the cylinder has circumference $2\pi$ and a finite height 
$h = \pi \, \frac{ \mathcal{K}(\sqrt{1-\zeta}) }{ \mathcal{K}( \sqrt{\zeta} ) } $,
where $\mathcal{K}$ 
is the complete elliptic integral of the first kind
and 
$\zeta = 
\frac{ (x_1 - x_2) \, (x_3 - x_4) }{ (x_1 - x_3) \,  (x_2 - x_4) }$  
is the cross ratio of the four points defining the geometry. 
The measurement slits are mapped to the boundaries of the cylinder, while the image of the subsystem $A$ is a horizontal line running along the surface (see End Matter for details of the map).

In turn, we have transformed our setup into a purely thermal one at inverse temperature $\beta_0=2\pi$, where the measurement outcome is encoded in the boundary conditions.
To preserve the conformal symmetry of our effective description, we shall require that the $U(1)$ (vector) current vanishes at the boundaries. This implies that the left/right-moving fermion modes on the cylinder differ only by a phase, $\psi_-(0,v) = \e^{\textrm{i}\alpha_1}\psi_+(0,v)$ and $\psi_-(h,v) = -\e^{\textrm{i}\alpha_2}\psi_+(h,v)$. Moreover, along the imaginary time direction $v$ we shall impose antiperiodic (NS) boundary conditions $\psi_\pm(u,v)=-\psi_\pm(u,v+\beta_0)$.

Our goal is now to evaluate the cylinder partition function.
Since one boundary phase can always be eliminated by a chiral transformation, the result depends only on the difference $\delta=\alpha_2-\alpha_1$. Furthermore, the cylinder geometry can be extended into a torus via the image method, i.e. identifying $\psi_-(u,v)=\psi_+(-u,v)$. One is then left with a chiral fermion on the torus, with a twisted BC along the spatial cycle of length $2h$. The effect of the twisted BC is to modify the quantization of the spatial modes, with allowed momenta being $q_k = (k+1/2-\nu)\pi/h$, where $\nu=\delta/(2\pi)$ and $k \in \mathbb{Z}$. Note that we defined the extra sign in the BC such that it becomes equivalent to the bosonized treatment in \cite{Khanna:2025fac}, with a zero mode appearing for $\delta=\pi$. The chiral torus partition function can then be computed as \cite{Alvarez-Gaume:1986rcs}
%
\begin{equation}
Z_\delta = \frac{\vartheta_3( \tau \nu | \tau)}{\eta(\tau)}\; \e^{\textrm{i}\pi\tau \nu^2},
\label{Zdelta}
\end{equation}
where $\vartheta_3$ is a Jacobi theta function 
\footnote{We use the convention $\vartheta_3(z|\tau)=\sum_{n=-\infty}^{\infty} \e^{\textrm{i}\pi\tau n^2} \e^{2\pi \textrm{i} n z}$} 
and $\eta$ is the Dedekind eta function
with modular parameter $\tau = \textrm{i} \tfrac{\beta_0}{2h}$. 
One can verify that the result is identical to the one obtained via bosonization in \cite{Khanna:2025fac,Khanna:2025mrz}. In particular, the exponential factor in \eqref{Zdelta} can be identified with the Born probability of the measurement outcome, described by a Gaussian distribution in terms of $\delta$.

At this point one should stress that the expression \eqref{Zdelta} is very reminiscent of a partition function in the presence of a chemical potential $\mu$. In fact, for (NS,NS) boundary conditions this reads \cite{Beneventano:2004zd,Ogawa:2011bz,Aguilera-Damia:2023jyc}
%
\begin{equation}
\label{Z-mu}
Z_\mu = \frac{\vartheta_3( \tau \frac{h \mu}{\pi} | \tau)}{\eta(\tau)}.    
\end{equation}
Comparing to \eqref{Zdelta}, one recognizes that the partition functions can be matched by setting $\mu=\mu_0=\delta/(2h)$, up to the exponential factor. Note that the latter corresponds to the Casimir energy term, which is modified by a twisted BC ($\delta\ne 0$) but left invariant by the chemical potential. However, they share the same essential feature of shifting the dispersion by a finite amount.

%
\begin{figure}[t]
\center
\includegraphics[width=\columnwidth]{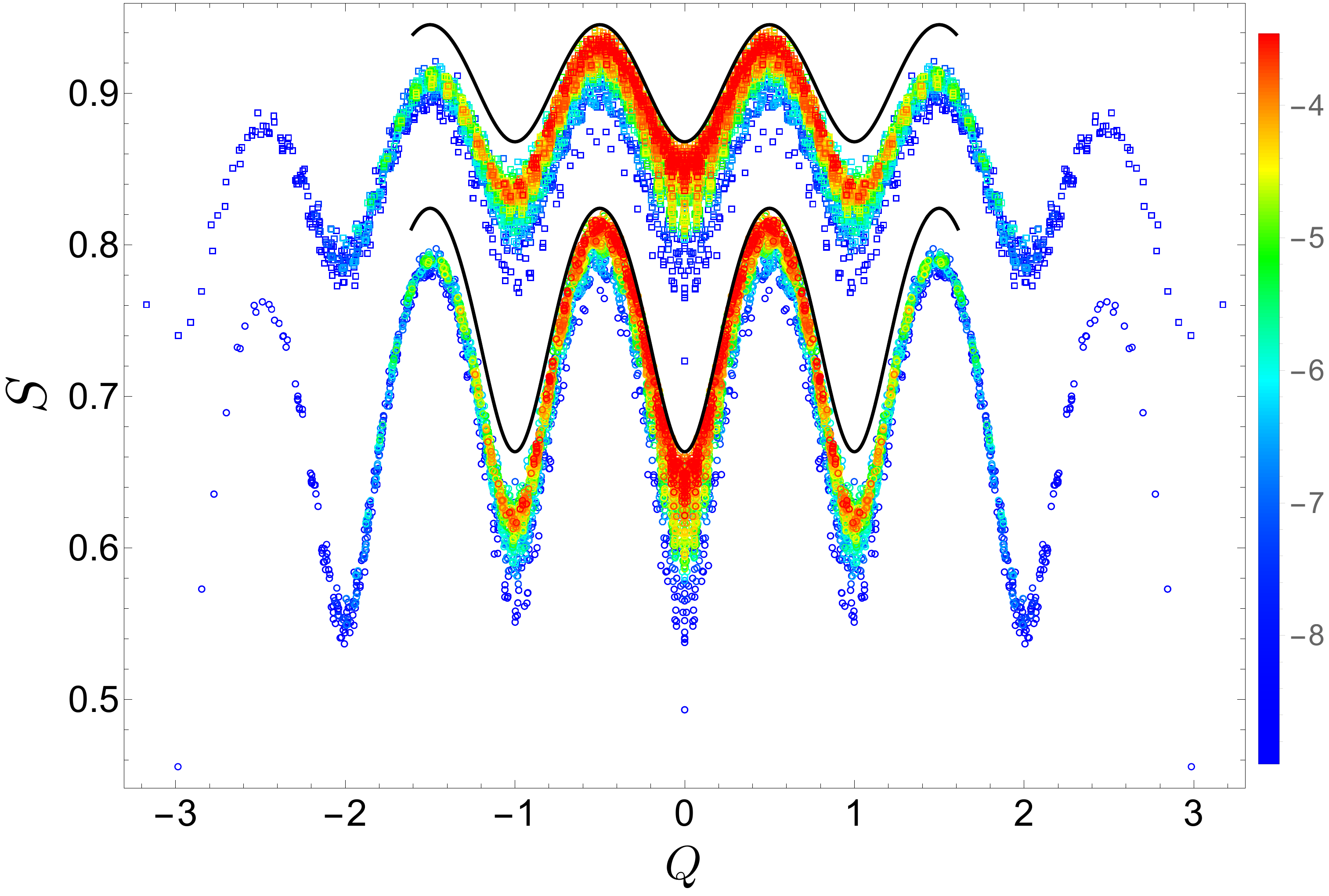}
\caption{Entanglement entropy against induced charge for all the $2^{2s}$ possible measurement outcomes, with $s=7$ and $\ell=20$ (top squares) as well as $\ell=10$ (bottom circles). The color code describes the corresponding Born probabilities on a logarithmic scale. The black solid lines show the CFT result, given by (\ref{Z-mu}) and \eqref{SQCFT}.}
\label{fig:SQ}
\end{figure}
%

Thus, in the CFT regime, the measurement-induced RDM is very closely related to a thermal system of effective size $h$
and with a chemical potential $\mu_0$.
In particular, they yield the same entanglement entropy as well as total charge within $A$, which are given by standard thermodynamical expressions as
\begin{equation}
S = \left(1-\beta_0 \frac{\partial}{\partial \beta_0}\right) \ln Z_{\mu_0}, 
\qquad
Q = \frac{1}{\beta_0}\frac{\partial}{\partial \mu_0} \ln Z_{\mu_0}.
\label{SQCFT}    
\end{equation}
The Casimir term in $\ln Z_\delta$ is linear in $\tau$
and therefore it  does not contribute in the expression of the thermal entropy. 
Hence, instead of the parameter $\delta$, the measurement-induced entanglement of a given outcome can be entirely characterized by the charge within $A$, described by a periodic function $S(Q)$ with period one.

The main question is, however, how well the CFT description can represent the measurement outcomes of the original lattice problem. For the free-fermion chain in \eqref{H} and in the occupation-number basis, the post-measurement state remains Gaussian, and its entropy can be computed via standard techniques based on the correlation matrix \cite{Peschel:2002yqj,Eisler:2009vye}, see End Matter for details. 
In Fig.~\ref{fig:SQ},
the entropy for all possible outcomes is plotted against the induced charge, relative to the unmeasured system. For simplicity, we consider a half-filled chain in a symmetric geometry, with $s = x_2-x_1=x_4-x_3$ and $\ell=x_3-x_2$, for two different values of the cross ratio. The corresponding data points form two bands, 
and their upper edge is remarkably well described by the CFT expression $S(Q)$, given by (\ref{Z-mu}) and (\ref{SQCFT}), in the regime $|Q|\lesssim 1.5$, as shown by the solid black lines. In fact, this edge represents the outcomes with the highest Born probabilities, as visualized by the color code. Data points further below the CFT curve occur with increasingly suppressed probabilities, representing outcomes that do not flow towards a conformal boundary condition. The decrease in probabilities for the larger $|Q|$ along the upper edge corresponds to the Gaussian weight in terms of $\delta$, and the deviation from the periodic CFT result is due to the effects of band-curvature on the lattice.


Although the above analogy gives a remarkably good characterization of the entropy in terms of the global charge, the local description of a twisted BC is different from that of a chemical potential. Indeed, while the latter gives a homogeneous charge density, for the former one expects the induced charge to be distributed close to the boundaries. This is immediately clear from the lattice realization of a conformal BC as a boundary chemical potential \cite{Eisler:2022rnp}, and also naturally dictated by the measurement mechanism. 
Finding an outcome of either $1$ or $0$ causes respectively 
either the charge deficit or the charge surplus 
(w.r.t.\ the average filling) 
to reorganize around the measured site, leading to an algebraic decay. The cumulated effect of
a bit string outcome in the segments $M_1$ and $M_2$ induces a nontrivial density, heavily peaked around the boundaries of $A$.

%
\begin{figure}[t]
\center
\includegraphics[width=\columnwidth]{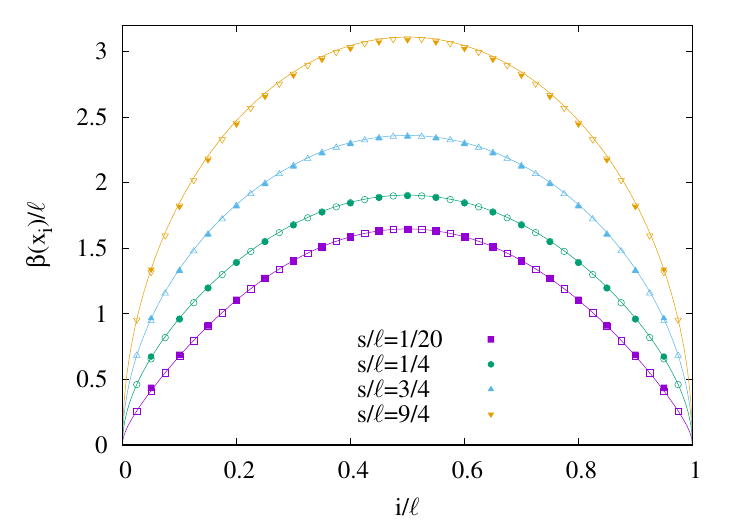}
\caption{Inverse temperature profiles $\beta(x)$ obtained from the continuum limit of the lattice EH,
for various $s$ and $\ell=20$ (filled symbols) as well as $\ell=40$ (empty symbols), with the ratio $s/\ell$ kept fixed.
The lines of matching color show the scaled CFT result \eqref{betax}.}
\label{fig:beta}
\end{figure}
%

These arguments imply that the RDM in the measurement setup is related to a grand-canonical state, with inverse temperature $\beta_0$ and chemical potential $\mu_0$, via the mapping $w(z)$ and an additional \emph{local} gauge transformation. Indeed, the latter can redistribute the charge density in $A$, while conserving the total charge. For the measurement-induced EH this suggests the ansatz
\begin{equation}
\mathcal{H}_\mathbf{m} =
\int_{0}^{\ell}  \beta(x)
\big[
T_{00}(x) - \mu_\mathbf{m}(x) \rho(x)
\big] \dd x \, ,
\label{HCFT}
\end{equation}
where $T_{00}(x)$ is the energy density of the massless Dirac fermion and $\rho(x)= \, :\mathrel{\psi_\textrm{\tiny R}^\dag(x)\psi_\textrm{\tiny R}(x) + \psi_\textrm{\tiny L}^\dag(x)\psi_\textrm{\tiny L}(x)}:$ is the normal ordered $U(1)$ density.
Note that it is coupled to the local chemical potential
$\mu_\mathbf{m}(x)$ that may depend on the \emph{total} measurement outcome, instead of the induced charge only.
In contrast, the local inverse temperature is expected to be $\beta(x)=\beta_0/w'(x)$, independent of the outcome $\mathbf{m}$, determined solely by the transformation properties of $T_{00}(x)$ under the conformal map \cite{Cardy:2016fqc}. 
As shown in the End Matter,
it takes the following simple form
\begin{equation}
\label{betax}
\beta(x)
= 
4  \,\mathcal{K}\big(\sqrt{\zeta} \,\big)\,
\sqrt{  \frac{ (x - x_1)\, (x - x_2)\, (x_3 - x)\, (x_4 - x)}{ 
(x_3 - x_1) (x_4 - x_2)}}\;.
\end{equation}
Remarkably, instead of the linear Bisognano-Wichmann 
scaling \cite{Bisognano:1975ih,Bisognano:1976za}, 
this local inverse temperature
follows a square root behavior 
close to the endpoints of $A$, 
and in the limit $\ell \to \infty$ 
with $M_1$ kept fixed one finds that 
$\beta(x) \to 2\pi \sqrt{(x - x_1)\, (x - x_2)}$. The prefactor $4  \,\mathcal{K}\big(\sqrt{\zeta} \,\big)\to 2\pi$ follows due to $\zeta \to 0$, which holds also in the limit $x_1\to x_2$ and $x_4 \to x_3$, where the expected parabolic profile of the unmeasured system is recovered
\cite{Hislop:1981uh, Casini:2011kv, Cardy:2016fqc}. It is worth considering also the asymmetric limit of (\ref{betax}) where e.g. only $M_1$ vanishes, 
finding that $\beta(x)$ has a linear (square root) behavior around the left (right) endpoint of $A$ \cite{Najafi:2016kwb}.
%

To check the CFT prediction for $\beta(x)$ on the lattice, we used standard free-fermion machinery to obtain the EH \cite{Eisler:2017cqi}, and performed the continuum limit of the resulting hopping matrix \cite{Eisler:2019rnr}.
This is formally based on the substitution 
$c_i 
\mapsto
\sqrt{a} \,\Big(  \e^{\textrm{i} q_F x_i}\, \psi_\textrm{\tiny R}(x_i) + \e^{-\textrm{i} q_F x_i}\,  \psi_\textrm{\tiny L}(x_i) \Big)$ 
between the lattice fermions and the left/right-moving Dirac fields at $x_i=a \, i$, followed by an expansion in terms of the lattice spacing $a \to 0$. We focus on half filling $q_F \, a=\pi/2$ and symmetric measurement intervals, and set $\ell \, a=1$, such that $A=[0,1]$. The results for the particular measurement outcome $\mathbf{m}_1=\dots1010$ and $\mathbf{m}_2=1010\dots$ (corresponding to induced charge $Q=0$) are shown in Fig.~\ref{fig:beta} for various values of $s$ and $\ell$. Despite the relatively small sizes, one finds an excellent agreement with the CFT prediction \eqref{betax}. For various other outcomes we find only tiny deviations in the numerical values $\beta(x_i)$, which are invisible on the scale of the figure.

%
\begin{figure}[t]
\center
\includegraphics[width=\columnwidth]{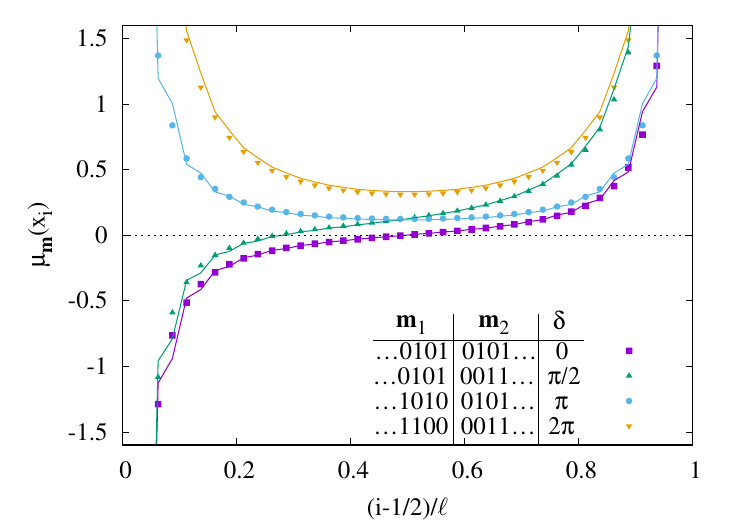}
\caption{Chemical potential $\mu_\mathbf{m}(x)$ obtained from the continuum limit of the lattice EH, for various measurement outcomes $\mathbf{m}=\mathbf{m}_1 \cup \mathbf{m}_2$ corresponding to different $\delta$ (symbols). The lines of matching color show the r.h.s. of Eq.~\eqref{murhoCFT}, with the density evaluated as in \eqref{rho} from the lattice data. The interval sizes are $\ell=40$ and $s=39$ for $\delta \in \{0,\pi\}$, 
as well as $s=38$ for $\delta \in \{\pi/2,2\pi\}$.}
\label{fig:mu}
\end{figure}
%

The continuum limit of the lattice results also allows for the identification of the chemical potential $\mu_\mathbf{m}(x)$.
Before presenting our numerics, however, we show how to fix this function on the level of CFT. As remarked earlier, this must be related to a local gauge transformation $g_\mathbf{m}(x)$ that, together with the conformal map $w(x)$, brings the EH \eqref{HCFT} to a grand-canonical form with homogeneous $\beta_0$ and $\mu_0$. Such a local gauge can be constructed using the 
transformation properties of $T_{00}(x)$ and $\rho(x)$ and yields \cite{Knizhnik:1984nr,Moosavi:2019fas,Mintchev:2025yso}
\begin{equation}
g_{\mathbf{m}}(x) = \int_0^x 
\left[ \mu_\mathbf{m}(y) - \mu_0 \frac{\beta_0}{\beta(y)} \right]  \dd y\,.
\label{gx}
\end{equation}
Note that the requirement to have a small gauge transformation (i.e. one that does not change the total charge) enforces 
$g_\mathbf{m}(\ell)=0$.

%
%


The final step is to relate $\mu_\mathbf{m}(x)$ to the charge density. To this end, we use the transformation property $\rho(x) \mapsto w'(x) \rho(w(x)) + g_\mathbf{m}'(x)/\pi$ under the combined conformal and gauge transformations, where the shift follows directly from the $U(1)$ current algebra. Taking the expectation values on both sides and using \eqref{gx} one finds
\begin{equation}
\mu_\mathbf{m}(x) =
\pi \braket{\rho(x)}_\mathbf{m} +
\frac{\pi}{\beta(x) h}(\delta - 2\pi Q) \, ,
\label{murhoCFT}    
\end{equation}
where we have substituted for $\mu_0$ and $\beta_0$ and used that the homogeneous density is $Q/h$. In turn, we see that the local chemical potential is proportional to the local density, plus a correction term that also involves $\beta(x)$.
Note that this second term vanishes at $\delta=Q=0$, as well as for $\delta=\pi$ where one has $Q=1/2$ for arbitrary $h$.

%
\begin{figure}[t]
\center
\includegraphics[width=\columnwidth]{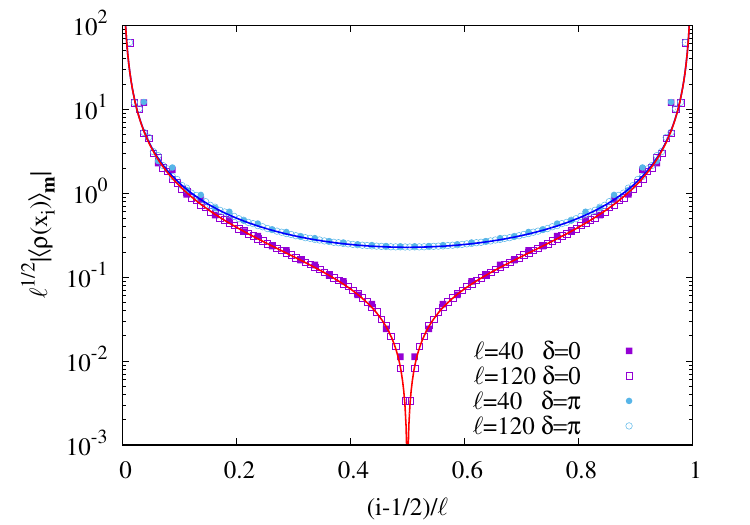}
\caption{Scaled charge density for the measurement outcomes with $\delta=0$ and $\delta=\pi$ (see Fig.~\ref{fig:mu}), for two different $\ell$ (filled/empty symbols) at fixed ratio $s/\ell=39/40$. The solid lines show the ansatz \eqref{rho}. The vertical scale is logarithmic.}
\label{fig:rho}
\end{figure}
%

In order to compare the result with the lattice data, we first need to properly match the corresponding density operators. Indeed, transforming to fields, the lattice density $n_i=c^\dag_ic_i$ acquires heavily oscillating terms with phases $\e^{\pm 2\textrm{i} q_F x_i}$ that mix the left/right-movers. At half filling, one can get rid of these alternating corrections by defining the induced density as
\begin{equation}
a \rho(x_i) = \frac{n_{i-1}+2n_{i}+n_{i+1}}{4} -\frac{1}{2} \, ,
\label{rho}    
\end{equation}
which holds up to corrections that vanish as $a\to 0$. Using this identification, one can compare $\mu_\mathbf{m}(x_i)$ obtained from the continuum limit of the EH \cite{Eisler:2019rnr} (see End Matter), to the r.h.s.~of \eqref{murhoCFT}. This is shown in Fig.~\ref{fig:mu} for various configurations, representing different $\delta$ values as summarized in the legend, with the dots indicating a periodic pattern.
One observes a very good agreement, with deviations increasing for larger $\delta$. Note that we omit the data points of largest magnitude close to the boundaries, where the continuum limit of the density in \eqref{rho} is not sufficiently smooth.

The EH thus carries information on the measurement outcome via the charge density, which is hard to evaluate in general. We find, however, that the simplest outcomes shown in Fig.~\ref{fig:mu} are extremely well described by the ansatz
\begin{equation}
\braket{\rho(x)}_\mathbf{m} = 
C a^{1/2} \left[\pm \, x^{-3/2} + (1-x)^{-3/2} \right],
\label{rho-bis}
\end{equation}
where the $\pm$ sign applies for $\delta=\pi$ and $0$, respectively. This is shown in Fig.~\ref{fig:rho} for two different $\ell$, with the ratio $s/\ell$ kept fixed. The fitted constant $C\approx0.04$ seems to be essentially independent of $\zeta$, similarly to the total charge. The density diverges close to the boundaries, but the integral regularized to $[a,1-a]$ gives a finite value. Note, however, that the continuous density \eqref{rho-bis} accounts only for a fraction of the total charge, while the rest remains localized at the boundary \cite{EislerTonni:tbp}.
%


In conclusion we have shown that, for partial measurements in the occupation basis, there is a clear relation between measurement-induced charge and entanglement entropy, at least for outcomes occurring with the highest probability.
Furthermore, the measurement qualitatively changes
$\beta(x)$ in the EH around the entangling points, and $\mu_{\mathbf{m}}(x)$ contains more detailed information about the charge density induced by a particular outcome. The locality of the EH should make it possible to test these results in quantum-simulator experiments based on EH tomography \cite{Kokail:2020opl,Joshi:2023rvd}. The method can naturally be extended to finite chain sizes or temperatures. Interactions could also be incorporated by considering a Luttinger liquid as in \cite{Khanna:2025fac,Khanna:2025mrz}, and it would be interesting to test how accurately the resulting local lattice EH could approximate the RDM \cite{Dalmonte:2017bzm,Giudici:2018izb}.
Finally, within the AdS/CFT correspondence, 
it would be insightful to find gravitational duals for  the quantities explored, e.g. in the AdS/BCFT setup discussed in \cite{Antonini:2022sfm}.
\\

We thank 
S. Murciano, 
M. A. Rajabpour,
D. Sexty, 
R. Vasseur 
and in particular M. Mintchev for fruitful discussions. This research was funded in part by the Austrian Science Fund (FWF) Grant-DOI: 10.55776/PAT3563424.


\bibliography{mieh_refs}

\onecolumngrid        

\newpage


\appendix
\section*{End Matter}

\section{Conformal map and local inverse temperature}
\label{EM-map}

The conformal map $z \mapsto w(z)$ employed in the main text 
(see the bottom part of Fig.~\ref{fig:map})
is the composition of three holomorphic transformations, through the sequence $z  \mapsto  \tilde{z} \mapsto \tilde{w} \mapsto w$.
The first holomorphic map $z  \mapsto  \tilde{z}(z)$  brings the measurement intervals in a configuration that is symmetric
with respect to the origin. 
Specifically, it takes the form 
\be
\label{z-tilde-def}
\tilde{z}(z) = \frac{a\, z + b }{c\, z + d}
\ee
where the real coefficients are obtained by imposing that 
\be
\label{tilde-z-constraints}
\tilde{z} (x_1) = - \frac{1}{k}
\; \qquad \; \tilde{z} (x_2) = - 1
\; \qquad \;  \tilde{z} (x_3) = 1
\; \qquad \;  \tilde{z} (x_4) = \frac{1}{k}
\ee
with $k \in (0,1)$ determined by the cross ratio 
$\zeta \in (0,1)$ of the four endpoints of the measurement intervals as follows
\be
\label{k-def-zeta}
k = \frac{1 - \sqrt{\zeta} }{ 1 + \sqrt{\zeta} }
\;\;\;\qquad\;\;\;
\zeta = \frac{ (x_1 - x_2) \, (x_3 - x_4) }{ (x_1 - x_3) \,  (x_2 - x_4) }.
\ee
The solution reads
\be
\label{abcd-parameters}
\begin{array}{l}
a = k\, (x_3 - x_2)  - (x_2 - x_1) - (x_3 - x_1)
\\
b = (x_3 - x_1) \, x_2 + (x_2 - x_1) \, x_3 - k\, (x_3 - x_2) \, x_1 
\\
c = k\, (x_2 - x_1 + x_3 - x_1) - ( x_3 - x_2 )
\\
d = (x_3 - x_2) \, x_1 - k \big[ (x_3 - x_1) \, x_2 + (x_2 - x_1) \, x_3   \big] .
\end{array}
\ee
The complex coordinate $\tilde{z} $ parametrises  the Riemann sphere $\mathbb{C}P^1$ 
with two slits of infinitesimal width along the measurement intervals 
$\widetilde{M}_1 = (-1/k,-1)$ and $\widetilde{M}_2 = (1, 1/k)$, whereas the subsystem $A$
is mapped onto $\widetilde{A} =(-1,1)$.

In the second step, this geometry is mapped onto an annulus 
with unit external radius.
The corresponding conformal transformation 
has been extensively discussed in Chapter\,VI of \cite{Nehari:book} and reads
%
\be
\label{tilde-w-map-def}
\tilde{w} (\tilde{z} ) = 
\exp \! \left[
\frac{h}{2} \left( \frac{ \textrm{arcsn}( \tilde{z} , k) }{ \mathcal{K}(k) } - 1\right)
\right]
\;\;\;\qquad\;\;\;
h  
=\, 
2\pi \, \frac{ \mathcal{K}(k) }{ \mathcal{K}( \sqrt{1-k^2} ) }
=\,
\pi \, \frac{ \mathcal{K}(\sqrt{1-\zeta}) }{ \mathcal{K}( \sqrt{\zeta} ) 
}
\ee
where $\textrm{arcsn}(\tilde{z}, k)$ is the inverse Jacobi elliptic function with modulus $k$ and $\mathcal{K}$ is the complete elliptic integral of the first kind.
The internal radius of the annulus is $\e^{-h}$, and \eqref{tilde-w-map-def} maps the segment
$\tilde{z}  \in (-1, 1)$ onto $\tilde{w}  \in (\e^{-h}, 1)$ on the real axis. The second form of $h$ in (\ref{tilde-w-map-def}) is obtained by applying the Landen transformations \cite{DLMF_Landen}
and tells us that the width of the annulus depends 
only on the cross ratio $\zeta$.

Finally, the third conformal map $w(\tilde{w} ) = \log(\tilde{w})$ sends the annulus onto the surface
of a cylinder of finite height $h$ and circumference $2\pi$, as depicted in the bottom part of Fig.~\ref{fig:map} in the main text. The composition of these three conformal mappings gives 
\be
\label{w-map-final}
w(z) =
\frac{h}{2} 
\left( \frac{ \textrm{arcsn} \! \big( \tilde{z}(z) , k \big) }{ \mathcal{K}(k) } - 1\right) .
\ee
This map satisfies $w(x_3) - w(x_2)  = h$, 
where $w(x_3) = 0$ .

The local inverse temperature $\beta(x)=2\pi/w'(x)$ is obtained through the derivative of (\ref{w-map-final}). 
This can be computed by using the elliptic integral representation
\be
\textrm{arcsn}(\tilde z,k)= \int_0^{\tilde z} \frac{\dd t}{\sqrt{(1-t^2)(1-k^2t^2)}}
\ee
which yields 
\be
\label{betax-EM}
\beta(x) =  
2\,\mathcal{K}\big( \sqrt{1-k^2} \, \big)
\frac{\sqrt{
\big(1-\tilde{z}(x)^2\big)
\big(1-k^2 \tilde{z}(x)^2\big)} 
}{ \tilde z'(x) }
=
4 \sqrt{k} \, \mathcal{K}\big( \sqrt{1-k^2} \, \big)\,
\sqrt{  \frac{ (x - x_1)\, (x - x_2)\, (x_3 - x)\, (x_4 - x)}{ 
(x_3 - x_2) (x_4 - x_1)}}  
\ee
where the last expression is found by 
using \eqref{abcd-parameters} in
the derivative of \eqref{z-tilde-def},
after some remarkable simplifications.
The expression in (\ref{betax-EM}) is 
equivalent to the one reported in Eq.\,(3.67) of \cite{Najafi:2016kwb},
which has a more complicated form.
Finally, the function of $k$ 
multiplying the square root in the last expression of 
(\ref{betax-EM}) can be written equivalently as 
$4\sqrt{k} \, \mathcal{K}\big(\sqrt{1-k^2} \,\big) 
= 
4\sqrt{1-\zeta}\, \mathcal{K}(\sqrt{\zeta}) 
=
2\pi \, \vartheta_4( 0 | - \!1/\tau)^2
=
2\pi \,(-\textrm{i} \tau) 
\,\vartheta_2( 0 | \tau)^2$,
where $\vartheta_2$ and $\vartheta_4$ are Jacobi theta functions
\footnote{
For these Jacobi theta functions, 
we adopt the convention  given by 
$\vartheta_2(z|\tau)=
\sum_{n=-\infty}^{\infty} \e^{\textrm{i}\pi\tau (n+1/2)^2} 
\e^{2\pi \textrm{i} (n+1/2) z}$ 
and
$\vartheta_4(z|\tau)=
\sum_{n=-\infty}^{\infty} (-1)^n \,\e^{\textrm{i}\pi\tau n^2} \e^{2\pi \textrm{i} n z}$, consistently with the one employed for $\vartheta_3(z|\tau)$.}
with $\tau = \textrm{i} \tfrac{\pi}{h}$, 
and the first identity of this sequence leads to the expression (\ref{betax}) 
in the main text.

\section{Correlation matrices and lattice EH}

Here we show how to construct the EH for the ground state $\ket{\psi_0}$ of the hopping chain, after a
particular outcome of the partial measurements. The ground state with Fermi momentum
$q_F$ is completely characterized by the correlation matrix, which is given by the sine kernel
\eq{
C_{ij} = \braket{c_i^\dag c_j} = \frac{\sin(q_F(i-j))}{\pi(i-j)}.
}
We now apply a projective measurement of the occupation number at a given site $m \in M$. This projects
the ground state into either $c^\dag_m c_m \ket{\psi_0}$ or $c_m c^\dag_m \ket{\psi_0}$, depending
on the measurement outcome being $1$ or $0$. Importantly, the projected state remains Gaussian,
and after a proper normalization by the corresponding Born probability $C_{mm}$, can still be completely
characterized by its correlation matrix. The matrix elements after the measurement can simply be evaluated
by Wick's theorem and read \cite{Cheng:2023bvw}
\eq{
(C_1)_{ij} =
\begin{cases}
1 & i=j=m \\
C_{ij} - \frac{C_{im}C_{mj}}{C_{mm}} \;\; & i \ne m, \, j\ne m\\
0 & \textrm{otherwise}
\end{cases},
\;\;\qquad\;\;
(C_0)_{ij} =
\begin{cases}
0 & i=j=m \\
C_{ij} + \frac{C_{im}C_{mj}}{1-C_{mm}}\;\; & i \ne m, \, j\ne m\\
0 & \textrm{otherwise}
\end{cases}
\label{Cupdate}}

The final state after measurements performed over the total domain $M=M_1 \cup M_2$ can simply be
obtained by applying the single-site update rule \eqref{Cupdate} sequentially. In turn, one arrives at
the matrix $C_\mathbf{m}$, where the vector $\mathbf{m}$ denotes the measurement record.
The corresponding Born probability $p_{\mathbf{m}}$ is easily evaluated as a determinant \cite{Rajabpour:2016fp}
\eq{
p_{\mathbf{m}} = \det P_{\mathbf{m}},
\qquad
(P_{\mathbf{m}})_{i,j} = 
\begin{cases}
C_{ij} & \textrm{if } \mathbf{m}_i=1 \\
\delta_{ij}-C_{ij} & \textrm{if } \mathbf{m}_i=0
\end{cases}, \quad i,j \in M
}
Furthermore, the entanglement entropy and charge induced by the measurement are given by
\be
\label{Sm-Qm}
S_\mathbf{m} = -\Tr [C_{\mathbf{m},A} \ln C_{\mathbf{m},A}+ (1-C_{\mathbf{m},A}) \ln (1-C_{\mathbf{m},A})], \qquad
Q_\mathbf{m} = \Tr \!\left[C_{\mathbf{m},A}- C_A \right]
\ee
where $C_{\mathbf{m},A}$ denotes the reduced correlation matrix, with indices restricted to the subsystem $i,j \in A$.
Note that $S_\mathbf{m}$ corresponds to the \emph{forced} measurement-induced entanglement, obtained after post-selection \cite{Rajabpour:2015uqa,Rajabpour:2015xkj,Najafi:2016kwb}, 
while the induced charge $Q_\mathbf{m}$ is measured w.r.t.\ the ground state. The symbols in Fig.~\ref{fig:SQ} correspond to \eqref{Sm-Qm} computed for each $\mathbf{m}$.

The EH pertaining to measurement outcome $\mathbf{m}$ is given by the standard free-fermion formula \cite{Peschel:2002yqj,Eisler:2009vye}
\eq{
\mathcal{H}_\mathbf{m} = \sum_{i,j \in A} (H_\mathbf{m})_{i,j} \, c^\dag_i c_j \, ,
\qquad
H_\mathbf{m} = \ln\!\left[ C^{-1}_{\mathbf{m}} -1 \right].
}
Note that the ground-state EH of a hopping chain is non-local,
i.e. the matrix $H_\mathbf{m}$ is not tridiagonal and contains longer-range hopping \cite{Eisler:2017cqi}.
Nevertheless, it is possible to remove these lattice effects by considering
a continuum limit, which allows for a proper comparison to CFT results.
In particular, using the substitution 
to Dirac fields reported in the main text,
and expanding in terms of the lattice spacing $a$, one finds at half filling $(q_Fa=\pi/2)$ the CFT expression \eqref{HCFT} with
the identification \cite{Eisler:2019rnr}
\eq{
\beta(x_i) = - 2a\sum_{p=0} (-1)^p (2p+1) (H_\mathbf{m})_{i-p,i+p+1} \, ,
\qquad
\beta(\tilde x_i)\mu_{\mathbf{m}}(\tilde x_i) = -(H_\mathbf{m})_{i,i} - 2\sum_{p=1} (-1)^p (H_\mathbf{m})_{i-p,i+p} \, ,
\label{betamu}}
where $x_i = i \, a$ and $\tilde x_i = (i-1/2)a$. Note that the arguments of the functions
associated to the kinetic energy and the density, respectively, are shifted by a half lattice spacing
for consistency with the lattice structure. The expressions in \eqref{betamu} were used to obtain the data points in Figs.~\ref{fig:beta} and \ref{fig:mu}.

\end{document}